\documentclass[]{spie}  
\usepackage{amsmath,amsfonts,amssymb}
\usepackage{graphicx}
\usepackage[colorlinks=true, allcolors=blue]{hyperref}
\usepackage{multirow}
\usepackage{upgreek}
\usepackage{gensymb}
\usepackage{mathtools} 
\usepackage[english]{babel}
\usepackage{lastpage}
\usepackage{hyperref}
\usepackage{rotating} 
\usepackage{caption, tabularx}
\usepackage{enumitem, ragged2e}
\usepackage[svgnames]{xcolor}
\usepackage{csquotes}
\usepackage{appendix}
\usepackage{epstopdf}
\usepackage{cancel}

\title{Low dosage 3D volume fluorescence microscopy imaging using compressive sensing}
\author[1]{Varun Mannam$^*$}
\author[2]{Jacob Brandt}
\author[2]{Cody J. Smith}
\author[1]{Scott Howard}
\affil[1]{Department of Electrical Engineering, University of Notre Dame, Notre Dame, IN 46556, USA}
\affil[2]{Department of Biological Sciences, University of Notre Dame, Notre Dame, IN 46556, USA}
\authorinfo{*e-mail: vmannam@nd.edu}
\begin{document} 
\maketitle
\begin{abstract} 
Fluorescence microscopy has been a significant tool to observe long-term imaging of embryos (\textit{in vivo}) growth over time. However, cumulative exposure is phototoxic to such sensitive live samples. While techniques like light-sheet fluorescence microscopy (LSFM) allows for reduced exposure, it is not well suited for deep imaging models. Other computational techniques are computationally expensive and often lack restoration quality. To address this challenge, one can use various low-dosage imaging techniques that are developed to achieve the 3D volume reconstruction using a few slices in the axial direction (z-axis); however, they often lack restoration quality. Also, acquiring dense images (with small steps) in the axial direction is computationally expensive. To address this challenge, we present a compressive sensing (CS) based approach to fully reconstruct 3D volumes with the same signal-to-noise ratio (SNR) with less than half of the excitation dosage. We present the theory and experimentally validate the approach. To demonstrate our technique, we capture a 3D volume of the RFP labeled neurons in the zebrafish embryo spinal cord (30 $\mu$m thickness) with the axial sampling of 0.1 $\mu$m using a confocal microscope. From the results, we observe the CS-based approach achieves accurate 3D volume reconstruction from less than 20$\%$ of the entire stack optical sections. The developed CS-based methodology in this work can be easily applied to other deep imaging modalities such as two-photon and light-sheet microscopy, where reducing sample photo-toxicity is a critical challenge.
\end{abstract}

\keywords{Low-dosage imaging, 3D volume reconstruction, fluorescence microscopy, compressive sensing (CS), discrete cosine transform (DCT), deep imaging, \textit{in vivo} imaging.}

\section{Introduction}
Understanding cellular biology during development 
requires long-term \textit{in vivo} imaging; however, many biological models are sensitive to photobleaching \cite{diaspro2006photobleaching} and phototoxicity where prolonged exposure to light can lead to tissue dysfunction or death \cite{reynaud2008light}. Accurate 3D volume reconstruction using confocal and multi-photon microscopy \cite{zhang2021instant}, however, requires dense imaging (with small steps) along the axial (z-axis) direction, leading to increased exposure and phototoxicity. Light-sheet fluorescence microscope (LSFM) \cite{reynaud2008light} is a commonly used imaging modality to achieve axial sectioning of tissue while reducing photodamage by illuminating only the required focal volume in the cross-section planes. While powerful, applications where delivering illumination light sheets deep in tissue can be prohibitively difficult for LSFM. Additionally, bandwidth and real-time processing of massive amount of data (typically in GB/TB) limitations required for acquisition and post-processing for the 3D volume reconstructions limit LSFM to applications everywhere \cite{reynaud2008light}. The image acquisition time is expensive to acquire dense images in the axial direction. With small number of axial images, 3D volume reconstruction suffers with poor signal-to-noise ratio (SNR) \cite{mannam2021real, mannam2020instant}. Several reconstruction methods are existing in the literature such as interpolation (linear, cubic-spline); however, they suffer from poor performance in reconstruction. For example, linear or cubic spline interpolation can reconstruct the slices from the missing z-axis slices for each pixel and repeat the same for all the pixels in the 2D image. However, interpolation-based methods are computationally expensive and highly ill-posed problem which cannot bring the missing information. Recently, machine learning (ML), particularly deep learning \cite{mannam2020machine}, has gained significant traction for its excellent performance in image processing such as finding the missing slices in a 3D volume, however, it requires huge training training data. There exists a couple of machine learning models that uses small datasets \cite{mannam2021deep, mannam2020performance}, but provides less accurate results. In this paper, we present a method to achieve 3D \textit{in vivo} imaging while minimizing optical dosages using a compressive sensing (CS) imaging technique. Through this technique, fewer axial frames are required to reconstruct a 3D volume while maintaining the same SNR as a complete stack.

\section{Methodology}
\begin{figure*}[!ht]
\centering
\includegraphics[page=1, width=1.0\linewidth]{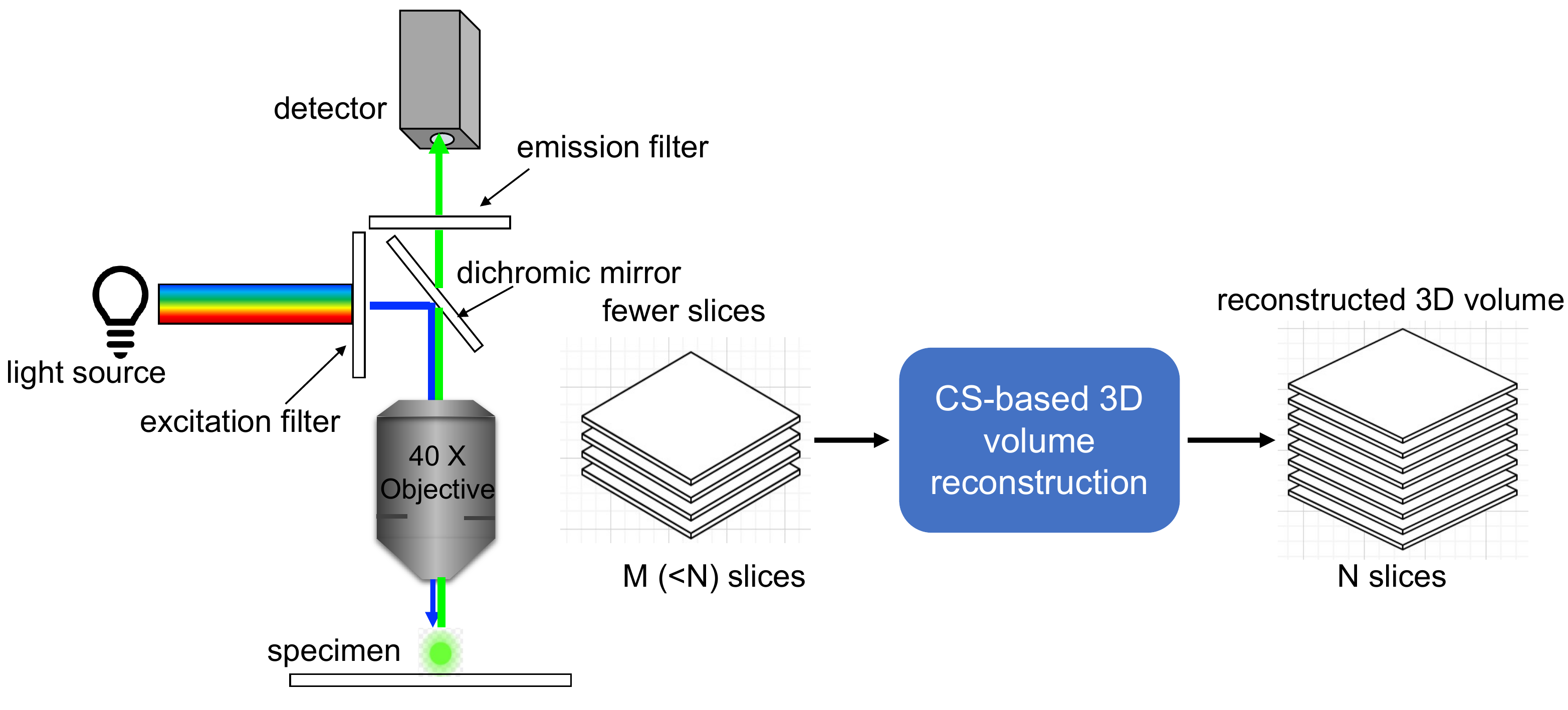}
\caption{Simple illustration of the demonstrated low dosage compressive sensing based 3D volume reconstruction approach.}\label{cs_fig1}
\end{figure*}

The data is collected for the following sections using a custom designed epi-fluorescence confocal microscopy, and setup is demonstrated in Fig.~\ref{cs_fig1}. An optical sectioning microscope (e.g., confocal, multi-photon, or LSFM) acquires optical sections along an axial plane (z-axis) while taking random steps (or skipping data excitation at random planes), then the entire volume is reconstructed using a compressive sensing based (CS-based) 3D volume reconstruction technique based on basis pursuit denoising \cite{chen1994basis, howard2020packet, marim2009compressed}. In this CS-based method, the complete 3D volume (at regular steps) can be reconstructed (estimated $N$ slices as output) from a fewer axial slices (using $M$ slices as input and $M<N$) than typical. The reconstructed CS 3D volume SNR is similar to original high dosage acquired 3D volume (acquired $N$ slices).

\begin{figure}[!ht]
\centering
\includegraphics[page=1, width=0.45\linewidth]{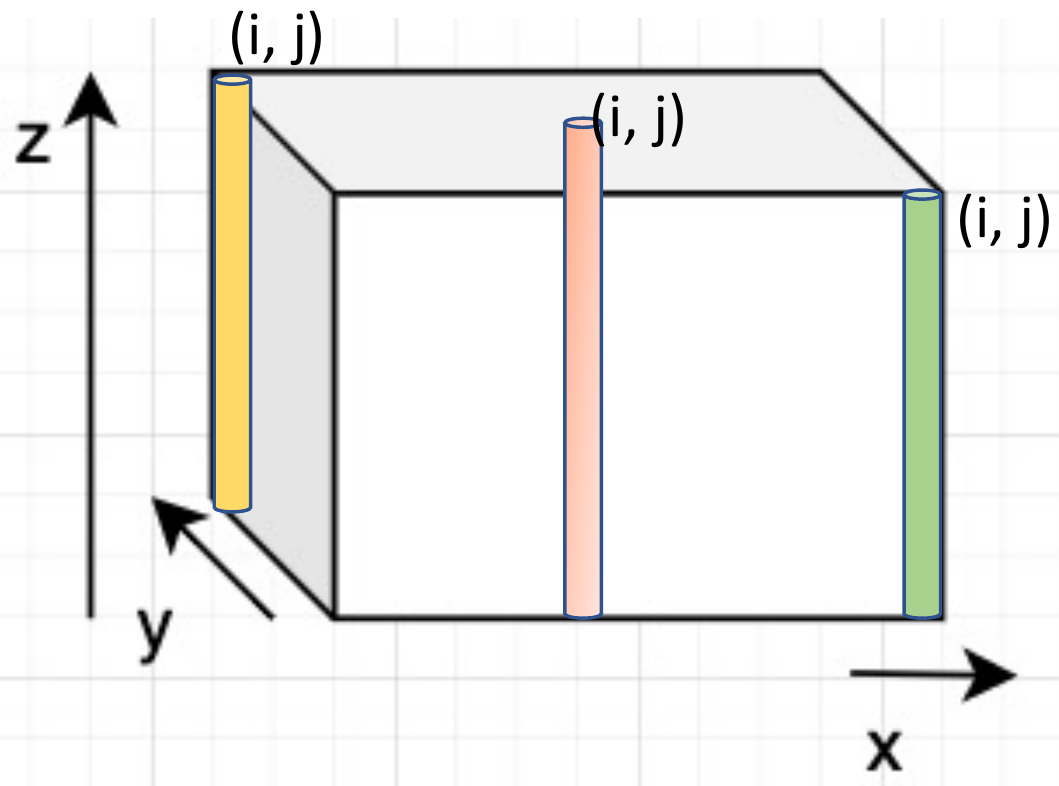}
\caption{3D volume representation indicating x and y including $z$-axis and selected 3 different pixels in different colors (start, middle, and end in the yellow, brown, and green, respectively) along the z-axis (1D data along z-axis for each pixel in the 2D position). Now along the $z$-axis, some of the slices are missing, and the CS-based method is to restore those missing pixels along the $z$-axis.}\label{cs2}
\end{figure}

Compressive sensing imaging exploits the fact that data can be represented in ``sparse" domains, i.e., domains where the complete image data is mostly zero, and therefore can be sampled sparsely within that domain without losing information. The grayscale image is filled with a lot of content, and it is not sparse in the spatial domain. However, this image is sparse in the transform domain. For example, the spatial frequency content of microscopy data is generally sparse with most of the non-zero values at lower spatial frequency components. In this case, we can perform a mathematical operation on an image called the discrete cosine transform (DCT), a real-domain transformation, unlike Fast Fourier transform (FFT), which is a complex domain transformation. The resulting ``DCT image" contains all the same information from the original image but is represented differently. In this transform domain, most of the biological samples are sparse under the condition that the image is over-sampled, which is valid for most of the cameras. Since the ``DCT image" is sparse and relatively simple to reconstruct the missing data (at missing z-axis). We can then do an inverse discrete cosine transform (IDCT) on the ``DCT image," and we will transform to get the reconstructed image accurately. In our approach, we performed a similar operation on the 1D values along the z-axis for each pixel to identify the missing slices in the 3D volume. Fig.~\ref{cs2} shows the 3D volume indicating the $x$, $y$, and $z$-axis, respectively, and slices are captured along the z-axis. A similar logic applies to the multi-channel images, including sparse representation in each channel (For example in case of color image, perform the method on individual R, G, or B channels). 

We will perform the DCT task computationally to find the DCT of z-axis values for each pixel such that:
\begin{enumerate}
    \item Sparse condition in the DCT domain.
    \item Estimate the DCT components such that reconstruction on the known slices (IDCT values) matches closely with the input values in the 1D data. 
\end{enumerate}

Mathematically, we are doing similar to ``3D basis pursuit denoising" to find the simplest values (i.e., the fewest non-zero DCT values) that match the pixels we have imaged so far. We can achieve this task by using a computer program (in python) to find the values of the DCT along the z-axis ($\mathbf{X}$) for a given pixel such that it gives you the smallest value of the following expression:

\begin{equation}
    \sum_m{ | \text{IDCT}(\mathbf{X})_m- b_m |^2} + C\sum{|\mathbf{X}|} \label{eq1}
\end{equation}
where $b_m$ is the value of the $m$th slice that was imaged for every pixel, and $C$ is a scaling factor (typically in the range of $3-5$). The first term is the sum of the squared error between the values in the reconstructed slices along the z-axis and the value of the slices that were actually imaged. Ideally, this will be zero. The second term is the L1 norm, which adds up all the values of $\mathbf{X}$, and minimizing that term is a suitable method for finding a sparse $\mathbf{X}$ \cite{andrew2007scalable}. Once you find $\mathbf{X}$, you can take the IDCT of $\mathbf{X}$ to see the reconstructed values along z-axis for a given pixel. Once the random set of optical sections have been acquired, the missing sections are computed by using the CS-based method on each pixel along the axial (z-axis) direction. For every pixel, we repeat the above mentioned method to find the missing values along z-axis. 

\textbf{Implementation}: Finding the DCT components for each pixel can be implemented using python in-built convex optimization libraries \cite{cvxpy}. Here the convex optimization code minimize the norm of the DCT components, with the constraint that reconstructed slices values in DCT should match up exactly with input random slices DCT components. However, this convex optimization is computationally expensive in time if the number of pixels are more in a given image.

To overcome this fundamental problem methods such as ``Limited-memory Broyden–Fletcher–Goldfarb–Shanno" (L-BFGS) algorithm \cite{lbfgs, lbfgs2} and its variant, the ``Orthant-wise limited-memory quasi-Newton" (OWL-QN) \cite{OWL, lbfgs3} are used. The OWL-QN algorithm allows to fit the Eq.~\ref{eq1} including minimizing the L1 norm of the DCT components. A simple demonstration of the reconstruction of the 1D data \cite{chen2001atomic} using the OWL-QN algorithm on the random sampling of the sinusoidal data is given \cite{rtpyrunner} and extension of this method on the 2D real images are provided in \cite{howard2020packet, pcsiwebsite}. 

\section{Results and discussion}
To demonstrate our approach, a 3D volume of red fluorescence protein (RFP) labeled zebra fish embryo (2 days post fertilization) spinal cord neurons was acquired using a spinning disk confocal microscope. The excitation wavelength is 400nm, emission wavelength of 488nm, the 3D volume thickness is 30 $\mu$m taken at 0.1 $\mu$m axial step resolution for a total of 301 slices in the axial direction. The stack is over-sampled with a small step size in axial-direction in order to perform the comparison study. Each 2D image dimension is 256$\times$256 pixels. From the complete stack, random sections could be removed to compare the CS reconstruction to the experimental data. We choose peak signal-to-noise ration (PSNR) and structural similarity index (SSIM) as the quantitative metrics to show the reconstructed 3D volume performance where the PSNR3D and SSIM3D are measured on the complete 3D volume. For the baseline comparison, the raw 3D volume (with 301 slices along z-axis) PSNR3D and SSIM3D are 37.6 dB and 0.94 with reference to the ground truth 3D volume. For reference, the complete stack averaged 10x was used as the ``ground truth" data.

\begin{figure*}[!ht]
\centering
\includegraphics[page=2, width=0.8\linewidth]{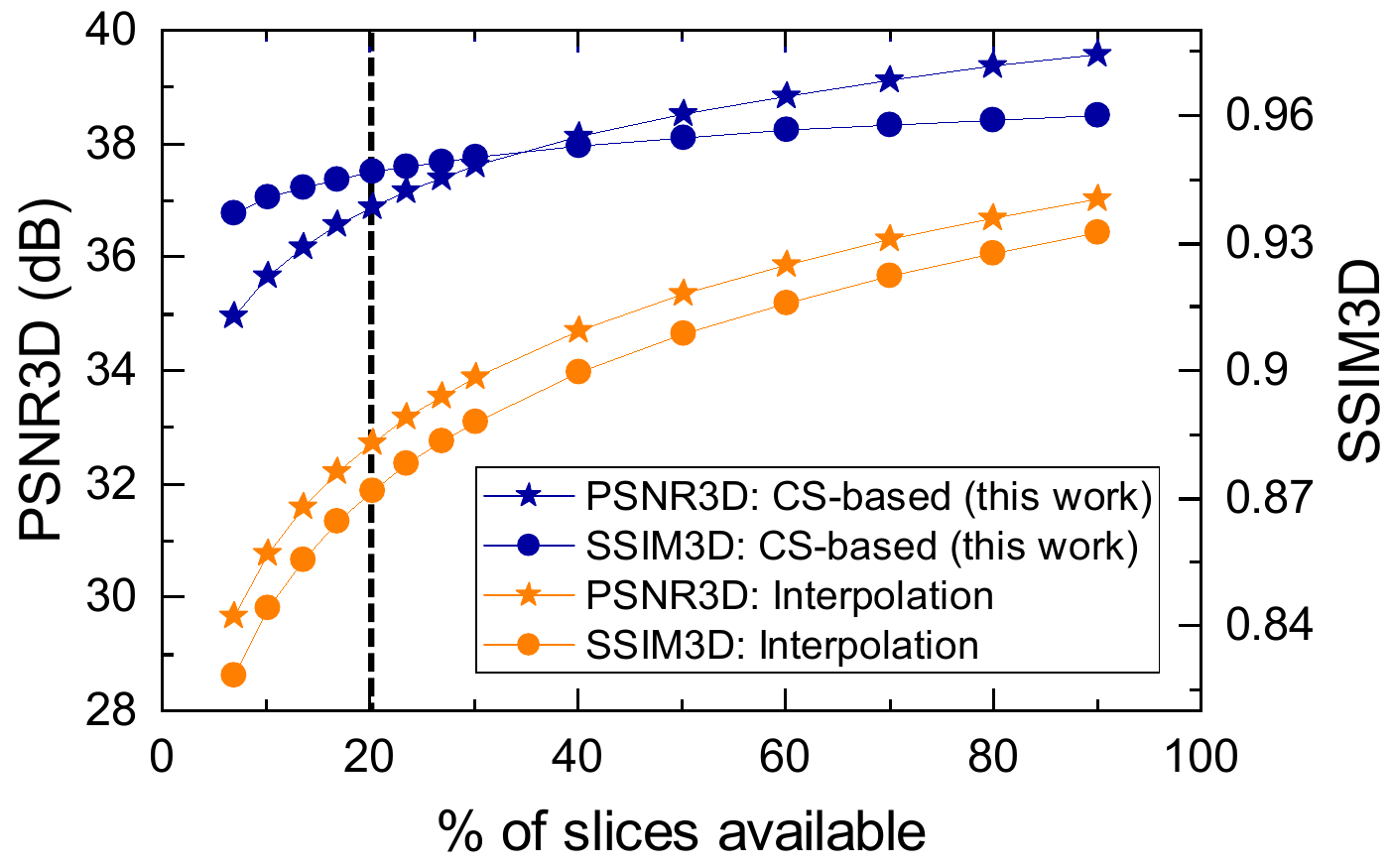}
\caption{Quantitative results of estimated slices using CS-based and interpolation methods. PSNR and SSIM of the reconstructed 3D volume show the cubic-spline and CS-based reconstruction methods on the estimated 3D volume from the missing slices (results are average PSNR3D from 100 iterations with random slices selected per iteration). Details of the sample and acquisition of the microscopy system are given in the text description.}\label{cs_fig2}
\end{figure*}
First, interpolation method is applied to estimate the missing slices (along z-axis) at every pixel when the input only a few slices (along z-axis) as indicated by the x-axis label (``\% of slices available"). Fig~\ref{cs_fig2} shows the quantitative comparison using the PSNR3D and SSIM3D metrics on the reconstructed volume with reference to the ground truth 3D volume. Fig~\ref{cs_fig2} orange plot shows the accuracy (in terms of PSNR3D and SSIM3D) of the interpolation method vs $\%$of slices available. Clearly, interpolation method underperforms when the $\%$of slices available is minimum. For example, for the $20\%$of slices available (shown in the gray dotted line in Fig~\ref{cs_fig2}), the PSNR3D and SSIM3D values are 32.73 dB and 0.87, respectively and these values are much smaller than the baseline metrics. Next, CS-based method is applied to estimate the missing slices (along z-axis) at every pixel to reconstruct the complete 3D volume. Fig~\ref{cs_fig2} blue plot shows the accuracy (in terms of PSNR3D and SSIM3D) of the CS-based method vs $\%$of slices available. From this plot, even with fewer slices available (for example: 20\%of slices available), quantitative metrics PSNR3D and SSIM3D are 36.88 dB and 0.95 respectively (better than interpolation method and baseline values. Also, the CS-based method outperforms the interpolation method for the complete 3D volume reconstruction with few slices available. This results shows the low-dosage 3D volume fluorescence microscopy imaging is possible using our demonstrated CS-based method. 

\begin{figure*}[!ht]
\centering
\includegraphics[page=3, width=1\linewidth]{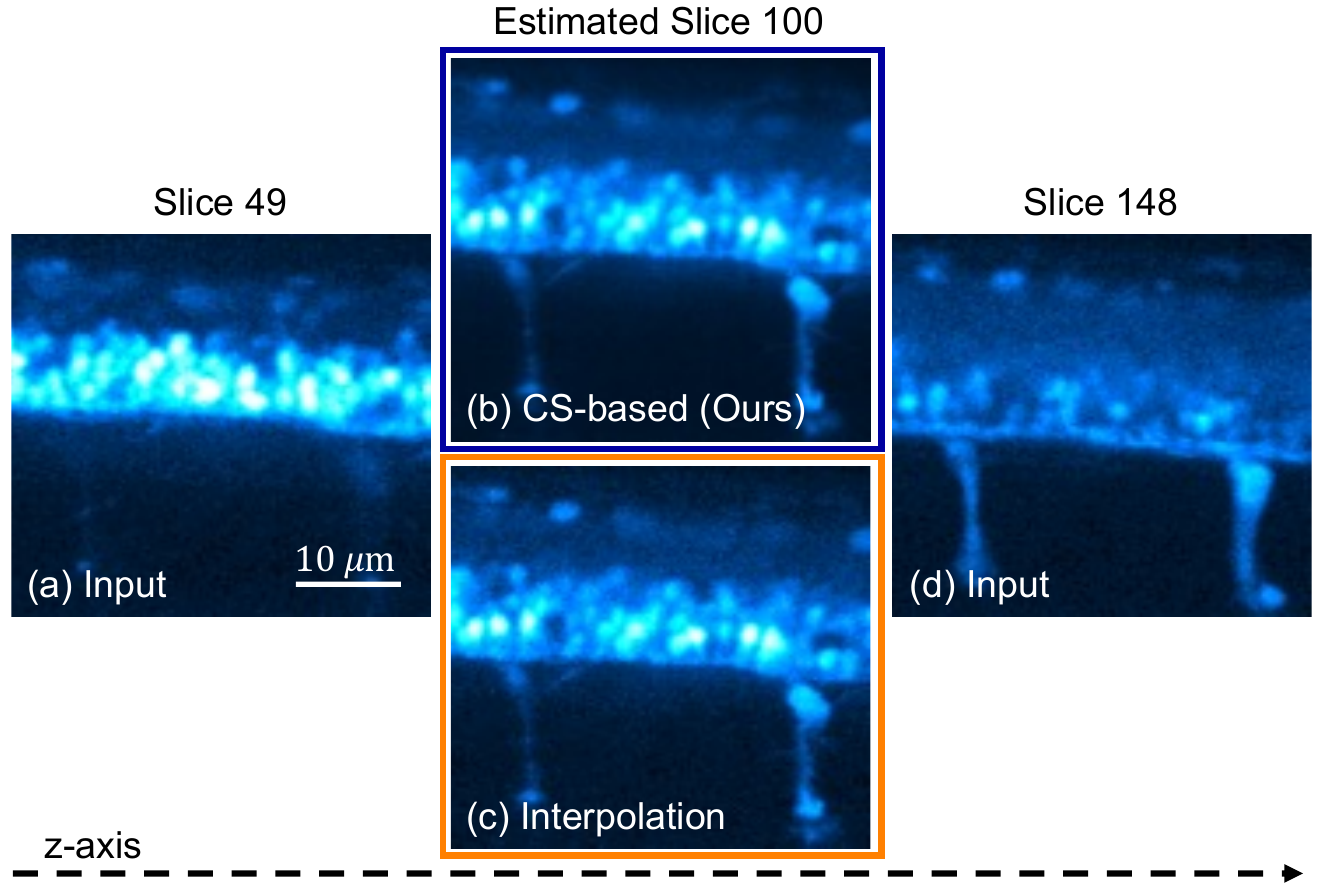}
\caption{Qualitative results demonstration of estimated slice (slice 100) in using the input slice 49 in (a) and slice 148 in (d) using our CS-based (b) and traditional interpolation (c) methods. Axial sampling is $0.1$ $\mu$m and slices of 49, 100 and 148 are corresponding to images at $4.9$ $\mu$m, $10.0$ $\mu$m and $14.8$ $\mu$m respectively. Sample and acquisition microscopy system details are given in the text description.}\label{cs_fig3}
\end{figure*}
Fig~\ref{cs_fig3} shows the qualitative results of the estimated slice from fewer input slices. Fig~\ref{cs_fig3}(b) and Fig~\ref{cs_fig3}(c) demonstrate the qualitative results of interpolation and CS-based approaches on missing slices (for example, estimation of $k^{th}$ slice from input of slices of $(k-1)^{th}$, $(k+1)^{th}$ as shown in Fig~\ref{cs_fig3}(a) and Fig~\ref{cs_fig3}(d) respectively of a zebra fish sample. From qualitative results, using CS-based approach is able to estimate $k^{th}$ slice which has improved PSNR compared to the interpolation-based estimated results. Using the CS-based approach (blue curves), an equivalent PSNR and SSIM to the complete whole stack can be achieved when removing approximately 50\% of the frames (i.e., reducing dosage in half). This method can be extended to provide 3D volume basis-pursuit denoising with limited dosage \cite{meiniel2018denoising} and can be extended to other image enhancement techniques such as super-resolution \cite{kim2021compressed} and deconvolution \cite{prakash2014basis}. Note: Here, the PSNR and SSIM are calculated on the complete 3D volume. The methods used in this manuscript are provided as open-source, accessible via GitHub\footnote{\url{https://github.com/ND-HowardGroup/Low-power-in-vivo-imaging.git}}.

\section{Conclusions}
Fluorescence microscopy in modern biology enables us to understand the embryo's growth over time using long-term \textit{in vivo} imaging; however, cumulative exposure is photo-toxic to live samples. Also, the accurate reconstruction of 3D volume requires dense (large number of) slices in the axial direction (z-axis), leading to increased exposure, and therefore leading to sample photodamage. Traditional methods like interpolation allow for 3D volume reconstruction using fewer slices but lack in obtaining good accuracy. We demonstrate a novel compressive sensing (CS) based approach is promising method to reconstruct the complete 3D volume using the low-dosage \textit{in vivo} imaging. The proposed method reduces the photodamage of the sample (zebrafish here) and thus enables long-term imaging. Our method shows with a minimum of 20$\%$ of available slices, can reconstruct the complete 3D volume when the input 3D volume is acquired with $0.1$ $\mu$m axial spacing. Also, the demonstrated method in this work looks promising where it can be integrated with other popular biomedical imaging modalities such as X-ray and MRI imaging. 

\section*{Disclosures}
\noindent The authors declare no conflicts of interest.

\section*{Funding.}
This material is based upon work supported by the National Science Foundation (NSF) under Grant No. CBET-1554516. 

\acknowledgments 
The authors further acknowledge the Notre Dame Center for Research Computing (CRC) for providing the Nvidia GeForce GTX 1080-Ti GPU resources for performing the compressive sensing data analysis in Python and MATLAB.

\bibliography{report} 
\bibliographystyle{spiebib} 

\end{document}